%% file: murphy_arxiv.tex
\documentclass[preprint]{aastex}

\makeatletter

 \newenvironment{lyxlist}[1]
   {\begin{list}{}
     {\settowidth{\labelwidth}{#1}
      \setlength{\leftmargin}{\labelwidth}
      \addtolength{\leftmargin}{\labelsep}
      }}
   {\end{list}}

\makeatother

\shorttitle{APOLLO: Lunar Laser Ranging}
\shortauthors{Murphy et al.}

\begin{document}

\title{APOLLO: the Apache Point Observatory Lunar Laser-ranging Operation:
Instrument Description and First Detections}

\author{T.\,W.~Murphy,~Jr.\altaffilmark{1},
E.\,G.~Adelberger\altaffilmark{2},
J.\,B.\,R.~Battat\altaffilmark{3},
L.\,N.~Carey\altaffilmark{4},
C.\,D.~Hoyle\altaffilmark{5},
P.~LeBlanc\altaffilmark{1},
E.\,L.~Michelsen\altaffilmark{1},
K.~Nordtvedt\altaffilmark{6},
A.\,E.~Orin\altaffilmark{1},
J.\,D.~Strasburg\altaffilmark{7},
C.\,W.~Stubbs\altaffilmark{8},
H.\,E.~Swanson\altaffilmark{2}, and
E.~Williams\altaffilmark{1}}
\email{tmurphy@physics.ucsd.edu}

\altaffiltext{1}{University of California, San Diego, Dept. of Physics, 9500 Gilman Drive, La Jolla, CA 92093-0424}
\altaffiltext{2}{University of Washington, Dept. of Physics, Seattle, WA 98195-1560}
\altaffiltext{3}{Harvard-Smithsonian Center for Astrophysics, 60 Garden Street, Cambridge, MA 02318}
\altaffiltext{4}{University of Washington Dept. of Astronomy, Seattle, WA 98195-1580}
\altaffiltext{5}{Humboldt State University, Dept. of Physics and Astronomy, One Harpst Street, Arcata, CA 95521-8299}
\altaffiltext{6}{Northwest Analysis, 118 Sourdough Ridge Road, Bozeman, MT 59715}
\altaffiltext{7}{Pacific Northwest National Labs, 902 Battelle Blvd., P.O. Box 999, Richland, WA 99352}
\altaffiltext{8}{Harvard University, Department of Physics, 17 Oxford Street,
Cambridge, MA 02318}

\begin{abstract}
A next-generation lunar laser ranging apparatus using the 3.5~m telescope
at the Apache Point Observatory in southern New Mexico has begun science
operation. APOLLO (the Apache Point Observatory Lunar Laser-ranging
Operation) has achieved \emph{one-millimeter} range precision to the
moon which should lead to approximately one-order-of-magnitude improvements
in several tests of fundamental properties of gravity.
We briefly motivate the scientific goals, and then give a detailed
discussion of the APOLLO instrumentation.
\end{abstract}

\keywords{Astronomical Instrumentation}

\section{Introduction}

\subsection{Scientific Motivation\label{sub:Scientific-Motivation}}

A variety of observations and theoretical explorations---including
the apparent acceleration of the expansion of the universe \citep{sn1,sn2},
the possible existence of extra dimensions \citep{extra_dim}, and
attempts to reconcile quantum mechanics and gravity---provide motivation
for improved tests of the fundamental aspects of gravity.

Lunar Laser Ranging (LLR) currently provides the best tests of a number
of gravitational phenomena \citep{jgw-96,jgw-latest} such as:

\begin{itemize}
\item the strong equivalence principle (SEP): $\eta\approx5\times10^{-4}$
sensitivity
\item time-rate-of-change of the gravitational constant: $\dot{G}/G<10^{-12}$
yr$^{-1}$
\item geodetic precession: 0.6\% precision confirmation
\item deviations from the $1/r^{2}$ force law: $\sim10^{-10}$ times the
strength of gravity at $10^{8}$ meter scales
\end{itemize}
LLR also tests other gravitational and mechanical phenomena, including
for example gravitomagnetism \citep{gravmag}, preferred frame effects
\citep{alpha-1,alpha-2}, and Newton's third law \citep{newtonsthird}.
LLR may also provide a window into the possible existence of extra-dimensions
via cosmological dilution of gravity \citep{lue,dvalimoon}. Besides
the SEP, LLR tests the weak equivalence principle (WEP) at the level
of $\Delta a/a<1.3\times10^{-13}$, but the LLR constraint is not
competitive with laboratory tests. In addition, LLR is used to define
coordinate systems, probe the lunar interior, and study geodynamics
\citep{dickey}.

These constraints on gravity are based on about 35 years of LLR data,
although the precision is dominated by the last $\sim$15 years of data at
1--3~cm precision.  APOLLO aims to improve tests of fundamental gravity by
approximately an order-of-magnitude by producing range points accurate at
the one-millimeter level.

\subsection{A Brief History of LLR\label{sub:A-Brief-History}}

The first accurate laser ranges to the moon followed the landing of
the first retroreflector array on the Apollo 11 mission by less than
two weeks (August 1, 1969). These were performed on the 3.0~meter
telescope at the Lick Observatory. One month later, a second station
using the 2.7~meter telescope at the McDonald Observatory began ranging
to the moon \citep{bender}. The operation at the Lick Observatory
was designed for demonstration of initial acquisition, so that the
scientifically relevant observations over the next decade came from
the McDonald station, which used a ruby laser with 4~ns pulse width,
firing at a repetition rate of about 0.3 Hz and $\sim3$ J/pulse.
This station routinely achieved 20~cm range precision, with a photon
return rate as high as 0.2 photons per pulse, or 0.06 photons per
second. A typical ``normal point''---a representative measurement
for a run typically lasting tens of minutes---was constructed from
approximately 20 photon returns.

In the mid 1980's, the McDonald operation was transferred to a dedicated
0.76~m telescope (also used for satellite laser ranging) with a
200~ps Nd:YAG laser operating at 10~Hz and 150 mJ/pulse. This station
is referred to as the McDonald Laser Ranging System: MLRS \citep{mlrs}.
At about the same time, a new station began operating in France at
the Observatoire de la C\^ote d'Azur (OCA) \citep{oca}. Using a 1.5~meter
telescope, a 70~ps Nd:YAG laser firing at 10~Hz and 75 mJ/pulse,
this became the premier lunar ranging station in the world. In recent
years, the MLRS and OCA stations have been the only contributors to
lunar range data with typical return rates of 0.002 and 0.01 photons
per pulse, respectively. Typical normal points from the two stations
consist of 15 and 40 photons, respectively.

Other efforts in LLR are described in \citet{ep-llr}, and more detailed
histories may be found in the preceding reference as well as in
\citet{bender,dickey}.

\subsection{Millimeter Requirements\label{sub:Millimeter-Requirements}}

The dominant source of random uncertainty in modern laser ranging
systems has little to do with the system components, but rather comes
from the varying orientation of the lunar retroreflector arrays. Although
the arrays are nominally pointed within a degree of the mean earth
position, variations in the lunar orientation---called libration---produce
misalignments as large as 10 degrees, and typically around 7 degrees.
This means the ranges between the earth and the individual array elements
typically have a root-mean-square (RMS) spread of 15--36~mm, corresponding
to about 100--240~ps of round-trip travel time. This dominates over
uncertainties associated with the laser pulse width, and with jitter
in the detector and timing electronics. A typical normal point containing
16 photons will therefore be limited to 4--9~mm range precision by
the array orientation alone, though range residuals reported by analysis
at the Jet Propulsion Laboratory tend to be larger than this.

Reaching the one-millimeter precision goal demands at a minimum the
collection of enough photons to achieve the appropriate statistical
reduction. Assuming an ability to identify the centroid of $N$
measurements---each with uncertainty $\sigma$---to a level of
$\sigma_{\mathrm{net}}=\sigma/\sqrt{N}$, the uncertainty stemming from the
retroreflector array orientation typically demands 225--1300 photons in the
normal point to reach the one millimeter mark. Worst-case orientations push
the individual photon uncertainty to 50~mm, demanding 2500 photons. This is
far outside of the capabilities of the aforementioned LLR stations. We
point out that any constant range bias is accommodated in the analysis, so
that only \emph{variations} in the range are important to the experiment.

While adequate photon number is sufficient to reduce statistical
uncertainty to the one-millimeter level, other sources of error could
potentially limit the ultimate scientific capacity of LLR.  Most
importantly, the gravitational physics is sensitive to the center-of-mass
separations of Earth and Moon, while one measures the distance between a
telescope and reflectors that are confined to the body surfaces.  The
earth's surface in particular has a rich dynamic---experiencing diurnal
solid-earth tides of 350~mm peak-to-peak amplitude, plus crustal loading
from oceans, atmosphere, and ground water that can be several millimeters
in amplitude.  Moreover, the earth atmosphere imposes a propagation delay
on the laser pulse, amounting to $\sim$1.5~m of zenith delay at
high-altitude sites.  Satellite laser ranging, very long baseline
interferometry, and other geodetic efforts must collectively contend with
these same issues, for which accurate models have been produced.  A good
summary of these models is published by the International Earth Rotation
and Reference Systems Service \citep[IERS:][]{iers}.

As an example of the state of these models, the long-standing
atmospheric model by \citet{marini-murray} has recently been replaced
by a more accurate model \citep{atmo1,atmo2}. The model differences
for a high-altitude site are no more than 2~mm for sky elevation
angles greater than 40 degrees---providing an indicative scale for the
model accuracy.  The primary input for this model is the atmospheric
pressure at the site, as this represents a vertical integration of
atmospheric density, which in turn is proportional to the deviation of
the refractive index, $n$, from unity.  Thus the zenith path delay,
being an integration of $n-1$ along the path, is proportional to
surface pressure under conditions of hydrostatic equilibrium.  A
mapping function translates zenith delay to delays for other sky
angles.  Measuring pressure to a part in 2000 (0.5~mbar) should
therefore be sufficient to characterize the 1.5~m zenith delay at the
one-millimeter level.  Our experiment records atmospheric pressure to
an accuracy of 0.1~mbar. 

The principal science signals from LLR appear at well-defined frequencies.
For example, the equivalence principle signal is at the synodic period of
29.53 days, and even secular effects ($\dot{G}$, precession) are seen via
the comparative phases between periodic (monthly) components in the lunar
orbit.  Because many of the effects discussed in the preceding paragraphs
are aperiodic, they will not mimic new physics.  To the extent that these
effects are not adequately modeled, they contribute either broadband noise
or discrete ``signals" at separable frequencies.

The science output from APOLLO may be initially limited by model
deficiencies.  But APOLLO's  substantial improvement in LLR precision,
together with a high data rate that facilitates deliberate tests of the
models, is likely to expose the nature of these deficiencies and therefore
propel model development---as has been historically true for the LLR
enterprise.  Ultimately, we plan to supplement our LLR measurement with 
site displacement measurements from a superconducting gravimeter (not yet
installed), in conjunction with a precision global positioning system
installation as part of the EarthScope Plate Boundary Observatory
(installed February 2007 as station P027).

\subsection{The APOLLO Contribution}

APOLLO---operating at the Apache Point Observatory (APO)---provides a major
improvement in lunar ranging capability. The combination of a 3.5~meter
aperture and 1.1~arcsecond median image quality near zenith translates to a
high photon return rate. Using a 90~ps FWHM (full-width at half-maximum)
Nd:YAG laser operating at 20~Hz and 115~mJ/pulse, APOLLO obtains photon
return rates approaching one photon per pulse, so that the requisite number
of photons for one-millimeter normal points may be collected on
few-minute timescales. To date, the best performance has been approximately
2500 return photons from the Apollo 15 array in a period of 8 minutes. The
average photon return rate for this period is about 0.25 photons per shot,
with peak rates of 0.6 photons per pulse. Approximately half of these
photons arrived in multi-photon bundles, the largest containing eight
photons. APOLLO brings LLR solidly into the multi-photon regime for the
first time.

This paper describes the physical implementation of the APOLLO apparatus,
including descriptions of the optical and mechanical design, the electronics
implementation, and system-level design. For early reports on APOLLO, see \citet{iwlr12,spacepart,iwlr13,iwlr14}.
For an analysis of our expected photon
return rate, see \citet{weak_LLR}. A list of acronyms commonly-used
in this paper appear in Appendix~\ref{app:acronyms}.

\section{System Overview\label{sec:System-Overview}}

\subsection{Overall Requirements and Differential Timing Scheme\label{sub:Requirements-Differential}}

Obtaining accurate laser ranges between a specific point on the earth's
surface and a specific point on the lunar surface requires that one
be able to establish both the departure and arrival times of the laser
pulse, referred to an accurate clock. One must also identify a spatial
location on the earth's surface to which the measurements are referenced.
For APOLLO, this corresponds to the intersection of azimuth and altitude
axes of the telescope, located near the tertiary mirror.
Light travels one millimeter in 3.3~ps, so that determining the one-way
lunar range to one-millimeter precision requires round-trip timing
that is accurate to the level of 6.7~ps. These considerations together
place stringent demands on the performance of the laser, clock, and
timing electronics.

Because the performance of the electronics can be a strong function of
temperature, and issues of mechanical flexure and thermal expansion become
relevant at the one-millimeter level, it is highly desirable to implement a
\emph{differential} measurement scheme.  As with other LLR stations, APOLLO
has a small corner cube in the exit path of the laser beam that intercepts
a small fraction of the outgoing pulse and directs it back toward the
receiver. These \emph{fiducial} photons follow the same optical path as the
lunar return photons---including all the same optical elements---but are
attenuated by a factor of $\sim10^{10}$ by thin reflective coatings that
are rotated into place by spinning optics. The fiducial return has a photon
intensity similar to that of the return from the moon, and is processed by
the detector and electronics in exactly the same manner as the lunar
return. By adjusting the fiducial rate to be between one and two photons
per pulse, it is possible to guarantee that a majority of outgoing pulses
have an associated fiducial measurement. 

Determining the start time via a single photon from a 90~ps full-width
at half-maximum (FWHM) laser pulse introduces an unnecessary uncertainty
in the measurement of the round-trip travel time. A higher signal-to-noise
ratio measurement can determine the laser fire time to higher
precision ($\sim15$~ps), though doing so disrupts the differential
nature of the measurement. It is, however, possible to accomplish
both goals at once: each laser pulse triggers a fast-photodiode (see
Section~\ref{sub:High-resolution-Timing}), producing a robust measurement
of the start time to 10--20~ps precision. Though the comparative
measurement of the start time as measured by the single-photon detector
and by the fast photodiode may may vary with time (as a function of
temperature, for instance), this variation is expected to be slow.
Thus the photodiode measurement can act as a timing ``anchor''
at a high signal-to-noise ratio for \emph{every shot}, and the single-photon
measurement of the start time can be used to ``calibrate'' the
timing anchor over few-minute timescales.

\subsection{Chief Components and Layout\label{sub:Layout}}

Figure~\ref{fig:overview} provides a schematic view of the APOLLO apparatus.
The APOLLO system consists of the following primary subsystems:

\begin{figure}
\begin{center}\includegraphics[%
  scale=0.75,
  angle=0]{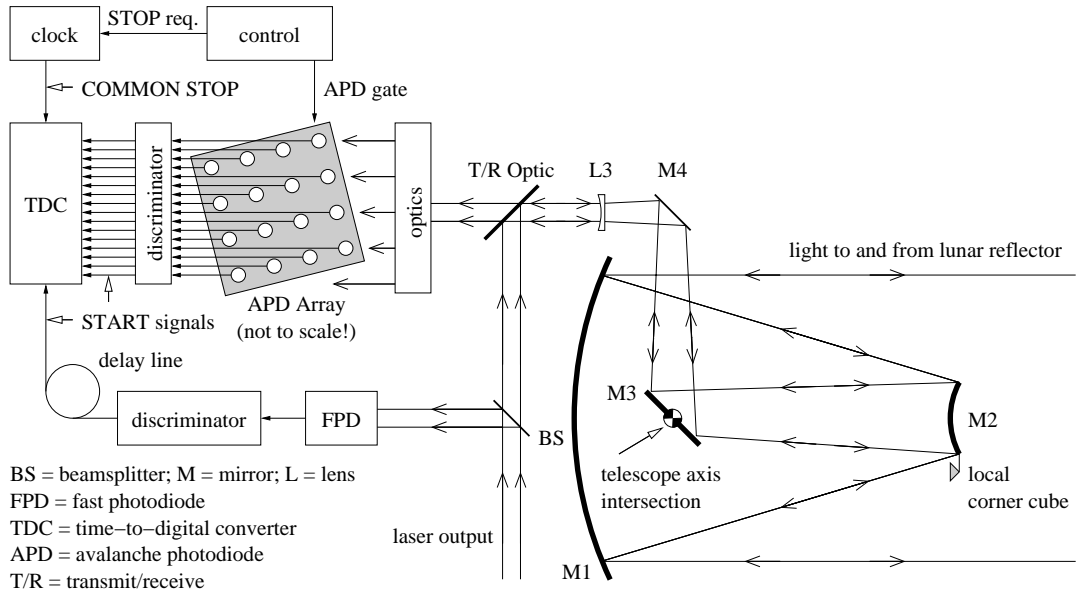}\end{center}

\caption{Overview of the chief components of the APOLLO system, showing the
telescope, local corner cube, detectors, and timing components.  Solid
arrowheads denote electronic signal paths, whereas open arrowheads denote
optical paths.  Optics labels follow the convention of the more detailed
optical layout in Figure~\ref{fig:optics}.  The box labeled ``control"
represents the hardware control computer as well as the Apollo Command
Module.  The box labeled ``clock" consists of the actual clock, clock
multiplier, and ``Clock Slicer" components.  The timing system components
are discussed in Section~\ref{sub:Timing-System}.\label{fig:overview}}
\end{figure}
\begin{itemize}
\item laser
\item optical system, including beam-switching optic
\item avalanche photodiode (APD) array detector
\item timing electronics (clock, counters, time-to-digital converter)
\item environmental monitoring and thermal regulation system
\end{itemize}
The laser, optical system, detector, and most of the timing electronics
are all affixed to the telescope. These components move with the telescope
and therefore experience a different gravity vector as the telescope
rotates about the elevation axis. The only discernible impact that
tilt has on our apparatus is a need to adjust the second-harmonic
generator crystal orientation in the laser, which is actuated.

\begin{figure}
\begin{center}\includegraphics[%
  scale=0.5,
  angle=0]{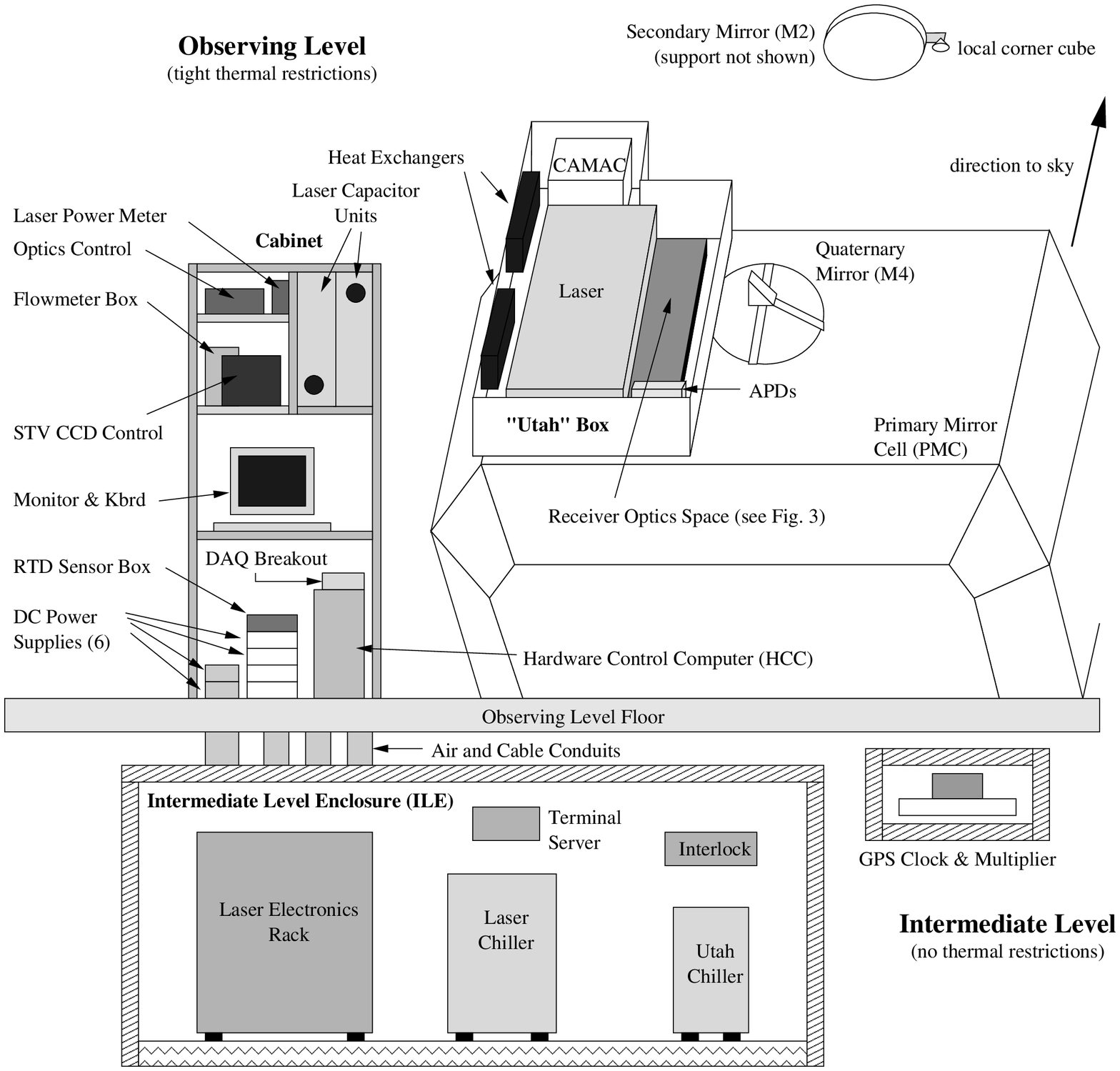}\end{center}

\caption{Schematic Layout of the APOLLO apparatus. The heavily-insulated
Utah-shaped box on the telescope contains the laser, APD detectors, and
timing electronics The insulated ``cabinet'' contains electronics that need
to be close to the telescope but not necessarily on the telescope.  The
large enclosure at bottom houses the high power-load equipment of APOLLO,
situated in an air volume separate from that of the telescope so that we
may dissipate heat into the surroundings. The GPS clock is also on this
level, in its own insulated box.\label{fig:layout}}
\end{figure}

Figure~\ref{fig:layout} shows the distribution of these pieces of
equipment. Four separate umbilical groups connect to the moving telescope.
Three of these groups run from the ``cabinet" to the laser enclosure.  Of
these three, one carries laser power and signals (including large cables
from the capacitor banks to the flashlamps); one carries signal cables for
device communications; and one carries de-ionized water to the laser heads
for cooling.  A fourth grouping that is well separated from the laser and
signal cables---connecting to the primary mirror cell near the elevation
axis---carries DC power and also propylene-glycol coolant for the ``Utah''
enclosure; More discussion of thermal control may be found in
Section~\ref{sec:Thermal-Control}.

\section{Optical Design\label{sec:Optical-Design}}

\begin{figure}
\includegraphics[%
  scale=0.65,
  angle=0]{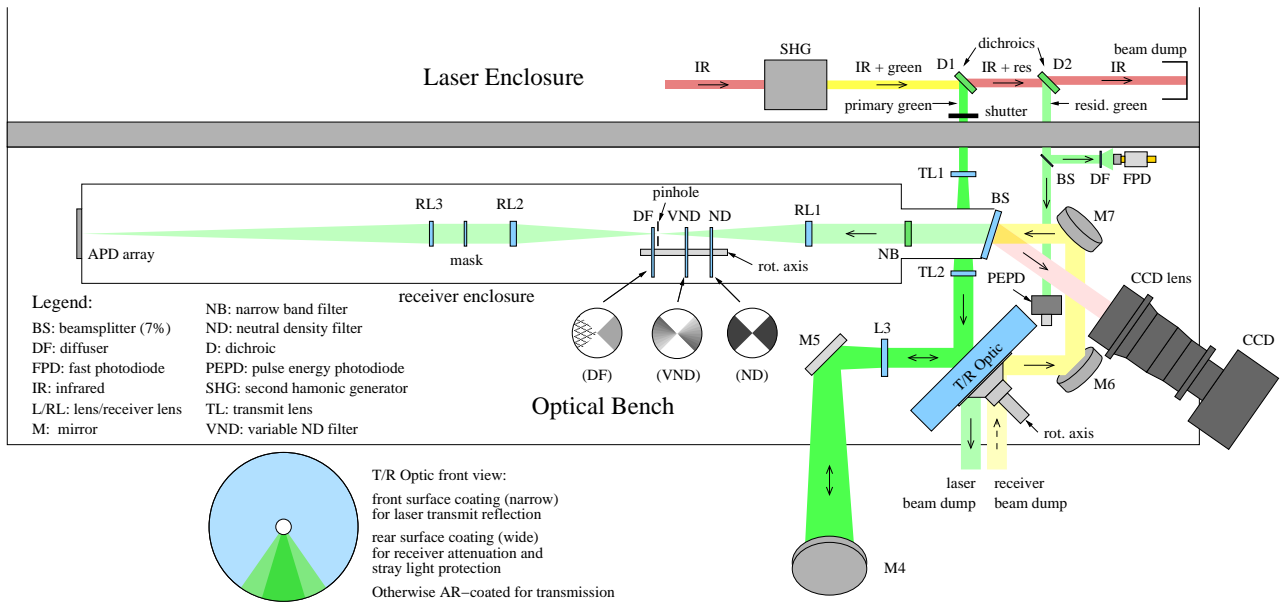}

\caption{The APOLLO optical layout. Components are labeled according to the
code table on the left. Beam directions are indicated. The T/R Optic
rotates at $\sim 20$~Hz, and the laser fires when the high-reflectivity
dielectric patch on the front surface rotates into position. The rear
coating on the T/R Optic, identical to the front coating except for
size, provides additional attenuation for both the local corner cube
return and the laser beam dump. The receiver ``beam dump'' provides
a dark target for the receiver during the laser flash, and coupling
between the laser beam dump and receiver ``dump'' is minimized.\label{fig:optics}}
\end{figure}

Figure~\ref{fig:optics} presents the layout of the APOLLO optical system.
The view is rotated $\sim90^{\circ}$ clockwise relative to the
orientation in Figure~\ref{fig:layout}. The following sections describe the
optical components.

\subsection{Laser and Laser Power Monitor\label{sub:Laser-and-Monitor}}

The laser is a Leopard solid-state picosecond product from Continuum
Lasers. Pumped by flashlamps at 20~Hz, 1.064~$\mu$m laser emission
from the Nd:YAG rod in the oscillator is shaped into $\approx120$~ps
pulses (FWHM) via the combined efforts of an acousto-optical mode-locker,
a solid state saturable absorber, and a GaAs wafer to clamp the pulse
energy. A double-pass amplifier boosts the cavity-dumped $~1$~mJ
pulse to $\sim250$~mJ, after which a second-harmonic-generator crystal
frequency-doubles the light at approximately 50\% efficiency to produce
115~mJ pulses of 532~nm light with pulse widths of about 90~ps
(FWHM).

Heat from the laser rods is carried away by de-ionized water flowing
at a rate of approximately 6 liters per minute, taking away $\sim1300$
Watts of thermal power. An auxiliary pump maintains flow in this loop
when the laser is not powered on and the ambient temperature is near
or below freezing.

Remote setting of the laser output power is provided by digitally
controlled potentiometers placed in the electronics units that control
the amplifier flashlamp delay and the voltage applied to the oscillator
flashlamp. The latter allow us to monitor the oscillator laser threshold
and adjust for optimal laser operation on a routine basis.

A single ``output'' dichroic (D1 in Figure~\ref{fig:optics}) sends the
green light out of the laser enclosure, past an output shutter actuated and
controlled by a custom interlock system. The infrared light passes through to a
beam dump. An actuated dichroic (not shown in Figure~\ref{fig:optics})
located before D1 alternately sends the green beam to a bolometric power
meter, so that we may periodically check the laser power. Another fixed
dichroic (D2) intercepts residual green light leaking through D1, sending
this to two detectors. The first is a photodiode (Hamamatsu S2281: labeled
PEPD in Figure~\ref{fig:optics}) that is used to integrate the pulse
energy, presenting the result as an analog output that is read and reset
after every pulse. The second is the fast photodiode (FPD) used in the
differential timing scheme. The function of this photodiode is twofold: (1)
establish a time reference of laser fire to $\sim15$~ps precision
(Section~\ref{sub:High-resolution-Timing}); and (2) alert the timing system
that the laser has fired, initiating the data acquisition cycle.

\subsection{Optical Train\label{sub:Optical-Train}}

The optical train (Figure~\ref{fig:optics}) has a transmit path
and a receive path that share the full aperture of the telescope.
The system must switch between transmit and receive modes (and back)
20 times per second. This is accomplished by the transmit/receive
optic (T/R optic). The T/R optic is a 30~mm-thick, 150~mm-diameter
glass disk anti-reflection (AR) coated (0.08\% reflection) for 532~nm
incident at 45$^{\circ}$ across most of the disk. A small sector
on the front of the disk has a multi-layer dielectric coating for
99.90\% reflection of S-polarized light at 532~nm. The disk is rotated
at about 20~Hz by a stepper-servo motor coupled via belt drive. The
angle encoder and index from the motor are processed by the Apollo
Command Module (Section~\ref{sub:ACM}), which directs the laser to
fire based on the position of the optic. The laser pulse---slaved
in this way to the T/R optic rotation---is arranged to strike
the highly reflective patch on the front of the T/R optic so that
it is sent to the telescope (transmit mode). Most of the time, the
T/R optic presents a clear path from the telescope aperture to the
receiver (receive mode).

\subsubsection{Transmit Path\label{sub:Transmit-Path}}

Starting at the laser output, the green pulse emerges as a $\sim7$~mm
diameter ($1/e^{2}$) beam with an approximately Gaussian profile,
centered 61~mm off of the optical bench. A bi-concave AR-coated lens
(TL1: $-74$~mm f.l.) followed by a plano-convex AR-coated lens (TL2:
168~mm f.l.)---both spherical---expand the beam to 16~mm diameter
prior to the T/R optic. Prior to June 2006, we used different lenses
that presented a 13~mm beam---and thus under-filled the telescope
aperture. Following the T/R optic, the beam encounters a plano-concave
lens (L3: $-198$~mm f.l.) that introduces a roughly $f/10$ divergence
to the beam so that it may nearly fill the telescope aperture. The
virtual focus of this lens is coincident with the telescope focus.
After L3, the beam experiences two 90$^{\circ}$ turns on M5 and M4---both
of which are multi-layer dielectric coatings for high-efficiency reflection
at 532~nm. After this are the telescope's aluminum-coated tertiary,
secondary, and primary mirrors (M3, M2, M1). The beam emerges from
the primary mirror collimated to well below 0.5~arcsec, as discussed
in Section~\ref{sub:Acquisition-and-Alignment}.

\subsubsection{Receiver Path\label{sub:Reciever-Path}}

Light from the telescope is brought toward a focus, following the
inverse path of the transmit beam, becoming collimated at L3. From
here, the path through the T/R optic experiences two 90$^{\circ}$
turns on M6 and M7, in the process being elevated to $\sim115$~mm
off of the optical bench so that it may cross the transmit path. M7
is tip-tilt actuated so that the receiver may be aligned relative
to the transmit beam direction (Section~\ref{sub:Acquisition-and-Alignment}).
The collimated beam enters the receiver tube via an uncoated glass
window, tilted to send the reflected light toward a charge-coupled
device (CCD) camera that aids acquisition and alignment. The clear
aperture up to this window is maintained to be at least 35~mm so
that a 40~arcsec field of view is preserved for the CCD camera. Past
this window, the optics are 25~mm in diameter, which is suitable
for the very small field of the avalanche photodiode detector array.

A narrow passband filter sits at the front of the receiver tube, with a
2.1~nm FWHM passband centered at 532~nm, and 95\% transmission at the
center wavelength. Prior to June 2007, we used a filter having 35\% peak
transmission and 1.5~nm bandpass. Beyond this, a doublet lens (RL1: 147~mm
f.l.) concentrates the collimated beam to a focus, where a pinhole is
placed to act as a spatial filter.  The 400~$\mu$m hole corresponds to
3~arcsec on the sky. An identical lens (RL2) is placed opposite the
pinhole, re-forming the collimated beam. An optional mask in the collimated
beam blocks light originating outside the telescope aperture,
but also serves an important purpose for the fiducial return (discussed in
Section~\ref{sub:The-Fiducial-Diffuser}).  A final lens (RL3: 347~mm f.l.)
focuses the light onto the detector at the end of the receiver tube. The
receiver tube is closely baffled at 50~mm intervals along its entire length
so that scattered light from the laser fire is unlikely to survive a trip
to the single-photon-sensitive detector.

\subsubsection{Detector Configuration and Microlens Array\label{sub:Lenslet}}

\begin{figure}
\begin{center}\includegraphics[%
  scale=0.6,
  angle=0]{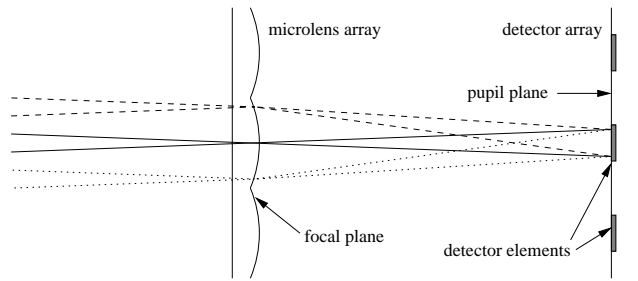}\end{center}

\caption{Microlens array, showing three field points as dashed, solid, and
dotted lines, respectively, within a single ``pixel'' of the lens
array. A pupil image is formed at the detector element.\label{fig:lenslet}}
\end{figure}

The detector (discussed in more detail in Section~\ref{sub:Detector-Array})
is a 4$\times$4 array device with 30~$\mu$m diameter active elements
in a square array on 100~$\mu$m centers. Thus the fill-factor is
low, at around 7\%. A microlens (lenslet) array is placed in front
of the detector so that the full fill-factor is recovered. The microlens
array is an epoxy replica on a 1~mm-thick glass substrate---each lens having
a 500~$\mu$m focal length and covering a 100~$\mu$m square. The
microlens array occupies the focal plane of RL3, where an image of
the far-field is formed. Each microlens element, or pixel, spans 0.35~arcsec
of angle on the sky, so that the focal plane is oversampled in typical
seeing conditions of 1.0~arcsec. At the detector, pupil images---effectively
images of the primary mirror---are formed on each 30~$\mu$m detector
element (Figure~\ref{fig:lenslet}). This means that the lunar return
photons illuminate an annulus on the detector with a central obstruction
due to the secondary mirror and an outer edge determined by the outer
radius of the primary mirror.

\subsubsection{Fiducial Attenuation\label{sub:Fiducial-Attenuation}}

An important ingredient in performing a differential measurement is
to establish identical conditions for the measurement of the lunar
photon returns and the fiducial returns from the local corner-cube.
This means: same optical path, same detector illumination, same signal
level, same electronics, etc. 

In order to achieve the same signal level, we must heavily attenuate the
light returning from the corner cube. Of the $3\times10^{17}$ photons in
the outgoing pulse, about $10^{13}$ strike the 25.4~mm-diameter corner
cube. This return has a diffractive spread that overfills the detector
array, and some attenuation is present in the form of narrow-band filter
throughput and detector efficiency (common to the lunar signal).  It is
nonetheless necessary to provide about ten orders-of-magnitude of
additional attenuation for the fiducial return.

Six orders-of-magnitude of attenuation---three from the front surface
coating and three from an identical rear-surface coating---are provided by
the T/R optic, which is virtually stationary during the $\sim 100$~ns
round-trip time to the corner cube. The thickness of the T/R optic is
determined by the requirement that reflections within the T/R optic do not
result in overlap between the primary beam and secondary ``ghost'' beams.
Also, the receiver, when ``looking'' off the reflective back side of the
T/R optic, should not see an overlapping patch of illumination from the
laser, since about $3\times10^{11}$ photons ($10^{-6}$ of pulse output) in
the outgoing pulse are transmitted through \emph{both} front and rear
reflective patches on the T/R optic. Well-engineered beam dumps with
$10^{5}$ suppression are placed side-by-side where the laser beam dumps and
the receiver ``looks'' to eliminate crosstalk between these paths (see
Figure~\ref{fig:optics}).

The remaining four orders-of-magnitude of necessary fiducial return
attenuation are achieved in the receiver tube near the spatial filter
pinhole. Here, a series of rotating disks with custom coatings present
an alternating pattern of attenuation and clear path to the fiducial
and lunar return photons, respectively. Two disks are placed before
the pinhole, and a third disk immediately behind. The choice to place
the disks here was dictated by the diffusing function, discussed in
Section~\ref{sub:The-Fiducial-Diffuser} below. The first disk alternates
between an optical density (OD) of 0.0 and 2.0. The second disk alternates
between ODs of 0.0 and 0.25--1.75. The range indicates that this optic
has an azimuthally varying (ramped) attenuation centered around OD
1.0, but spanning $\pm0.75$ OD on either side. This ramp allows tuning
of the fiducial attenuation by setting the phase of the rotating disks
relative to the T/R optic. The final disk in the series alternates
between OD 0.0 and 1.15---rounding out the targeted total of OD $\sim4$.
The rotation of the disks is accomplished by a stepper motor slaved
to the T/R optic motor, so that relative phasing is stable and controllable.

\subsubsection{The Fiducial Diffuser\label{sub:The-Fiducial-Diffuser}}

Each of the three disks in the arrangement described above is coated
in quadrants and spun at half the speed of the T/R optic (i.e., around
10~Hz). Opposite quadrants are either clear (AR-coated) or attenuating,
giving a bow-tie appearance to each disk, seen in the inset figures
in Figure~\ref{fig:optics}. The third disk is different from the
first two in that opposite the OD 1.15 attenuator is not an identical
attenuator but a quadrant of ground glass to act as a diffuser. The
purpose of the diffuser is to spread the fiducial photons so that
the detector is illuminated in exactly the same manner as it is by
the lunar photons.

As described in Section~\ref{sub:Lenslet}, an image of the primary
mirror is formed on each of the detector elements. Because the response
time of the detector depends on the position of the incident photon,
a true differential measurement requires identical spatial illumination of
the detector in both circumstances. But the fiducial photons all come
from a small corner cube located near the secondary mirror. The image
of this corner cube on the detector would then be small, looking nothing
like the illumination pattern from the lunar photons. The diffuser
located adjacent to the spatial filter pinhole spreads the fiducial
photons out, affecting a uniform illumination of the detector element.
The optional mask between RL2 and RL3 (not yet installed) would allow
one to impose precisely the same illumination pattern on both lunar
and diffused fiducial returns, complete with the central obstruction.
See Section~\ref{sub:Detector-Array} for an estimate of the effect the
mask would have.

The spatial dependence of the detector response time means that one
sacrifices temporal clarity when diffusing the fiducial photons. Any
comparative bias has been removed, but at the expense of a temporal
spread.  Taking full advantage of the timing system to characterize
the laser pulse width and detector response, the quadrant opposite the
diffuser is a simple attenuator, so that the illumination of the
detector is still confined, and thus suffers less temporal spread.  In
this way, it is still possible to carry out system diagnoses with high
temporal resolution. Additionally, one may directly measure the
detector's temporal bias by comparing the time offset of each type of
fiducial return (diffuser vs. attenuator) against the fast photodiode
timing ``anchor'' for each pulse.  In practice, we find the average
offset between the two to be about 10~ps.

\subsection{Detector Array\label{sub:Detector-Array}}

\begin{figure}
\begin{center}\includegraphics[%
  scale=0.6]{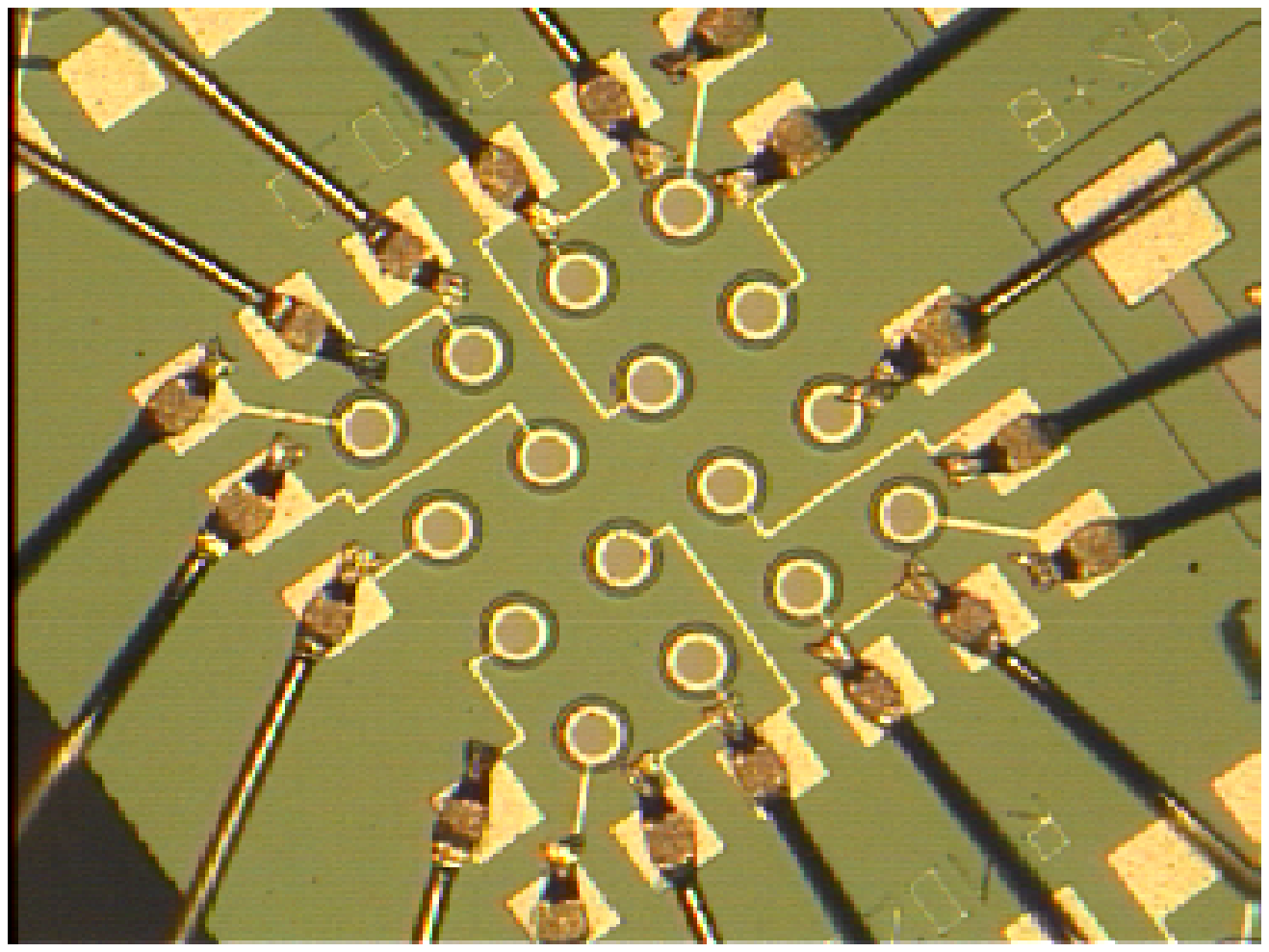}\end{center}

\caption{The 4$\times$4 APD array, with 30~$\mu$m elements on 100~$\mu$m
centers.\label{fig:APD}}
\end{figure}

The APOLLO detector is a 4$\times$4 avalanche photodiode array fabricated
at Lincoln Lab (Figure~\ref{fig:APD}) \citep{jana_spie}. The elements
are 30~$\mu$m in diameter on 100~$\mu$m centers in a square pattern.
The material is lightly-doped \emph{p}-type silicon, epitaxially grown
onto a $p^{+}$ substrate. A buried $p^{+}$ layer is implanted about
1.0~$\mu$m deep, acting as the multiplication region. The front
surface is heavily doped via diffusion to make an $n^{+}$ layer and
thus the \emph{p}-\emph{n} junction. The breakdown voltage is about
25~V, and a typical bias voltage is about 30~V, placing the device
in avalanche Geiger mode. The detector spends most of its time biased
just below the breakdown voltage, and biased above breakdown for only
a 180~ns window, or gate, around the time of an expected photon
arrival.  The timing window spans 100~ns within the larger APD window.

The device is front-illuminated, so that most photoelectrons arising
from 532~nm photons---with a penetration depth of about 1.0~$\mu$m---very
quickly avalanche without spending time in a slow diffusion state.
The depletion region in the overbiased state extends a few microns
into the device, meaning that a driving electric field exists in this
volume. Photoelectrons created within the depletion region are driven
at the saturation velocity of 0.1~$\mu$m/ps toward the avalanche
region. Photons that happen to penetrate below the depletion region
generate photoelectrons that may wander for tens of nanoseconds before
encountering the depletion region, after which the avalanche is prompt.

One aspect of the avalanche process that impacts our measurement is
the lateral propagation of the avalanche microplasma within the thin,
disk-shaped multiplication region. Starting in a spot close to where
the photon enters the detector, the boundary of this microplasma
expands in a roughly linear fashion at a speed close to the thermal
velocity of electrons in silicon. Since the avalanche current is proportional
to the area of the avalanching region, the current grows quadratically
in time until one edge of the microplasma encounters an edge of the
multiplication region---or active region of the detector element.
Ultimately, full avalanche current is reached when the entire multiplication
region is participating in the avalanche. Depending on the current
level to which the trigger electronics are sensitive, this phenomenon
could result in a spatially-dependent delay of trigger. If the electronics
are capable of triggering at a low threshold (but therefore more susceptible
to noise triggers), the avalanche does not have to be very large to
trigger, and the delay between photon arrival and electronic trigger
will be relatively insensitive to the location of the incident photon
until this location approaches the edge of the multiplication region.
If, on the other hand, the trigger level is set to a high value such
that the majority of the active area must be in avalanche, a photon
landing in the center will achieve this state long before a photon
impinging near the outer edge---by as much as a few hundred picoseconds
\citep{jana_thesis}. The intermediate case would see time-delay insensitivity
for a circular region around the center, but acquiring a spatial dependence
outside of this region. It is for this reason that APOLLO employs
a diffuser for the corner cube fiducial returns, as discussed in Section~\ref{sub:The-Fiducial-Diffuser}.
Calculations indicate a reduction of bias from about 100~ps to about
5~ps when a diffuser is employed for high trigger levels.

Because we set the avalanche trigger at a low level---about 20~mV
compared to a $\sim 120$~mV full avalanche---the spatial dependence of
the temporal response is somewhat suppressed.  We estimate the effect
that the optional mask would have on the timing to be about 1.7~ps,
based on the 10~ps scale of the difference between diffused and
non-diffused fiducial timing.  It should be noted that while attempts
to reduce known sources of bias in the differential measurement are
warranted, \emph{static} biases will not compromise APOLLO's science
goals.

Crosstalk has been extensively characterized for these arrays. The
energetic avalanche process occasionally gives rise to photon emission,
which can propagate through the silicon to be absorbed in the active
region of a nearby otherwise quiescent element, causing it to undergo
avalanche. A simple model of this phenomenon suggests a $1/r^{4}$
probability for crosstalk, where $r$ is the separation between elements.
This relationship is confirmed by experiment. Nearest neighbors (100~$\mu$m
separation) experience a steady-state crosstalk rate of 0.001 events
per nanosecond while the avalanching element is held in a full avalanche
current regime. Therefore, a photon-induced avalanche midway into
the $\sim100$~ns gate results in a 5\% avalanche probability due
to crosstalk in each nearest-neighbor, down by a factor of four in
the next-nearest neighbors.

One surprising aspect of the crosstalk phenomenon is that it only
slowly reaches a steady state rate. If we cause element A to avalanche,
and sustain the avalanche current for $> 100$~ns, while monitoring element
B for a correlated event, we find that the crosstalk rate exponentially
approaches its steady-state value with a time constant of about 30~ns.
There is no sign of prompt crosstalk avalanche events. Practically,
this means that since all of the lunar photons return within 1~ns
of each other, the crosstalk phenomenon is not yet operative on this
timescale so that we do not have to worry about crosstalk events masquerading
as genuine lunar returns.

\subsection{Acquisition and Alignment\label{sub:Acquisition-and-Alignment}}

\begin{figure}
\begin{center}\includegraphics[%
  angle=0]{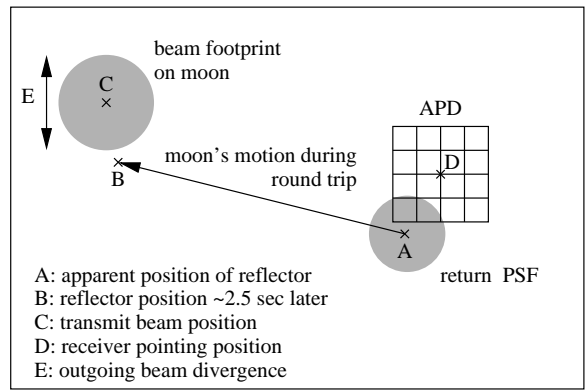}\end{center}

\caption{Depiction of pointing parameters. The lunar return is seen at position
A, where the moon was 1.25 seconds ago, while the outgoing laser beam
must point at B, where the moon \emph{will} be in 1.25 seconds. Meanwhile,
the receiver (APD) is pointed imperfectly at D, and the laser is
striking off-target at B. The size of the return point spread function
(PSF) is determined by atmospheric seeing, while the footprint of
the beam on the moon is governed both by seeing and improper focus/collimation.
Ideally, D is coincident with A, C is coincident with B, and the lunar
footprint is no larger than the seeing limit.\label{fig:pointing}}
\end{figure}

Closing the laser link to the lunar reflectors is a demanding task,
requiring simultaneous satisfaction of six variable parameters. One
is the outgoing beam divergence, closely related to telescope focus.
One is the timing of APD activation---a $\sim100$~ns window that
must be precisely positioned in time. Two relate to the telescope
pointing, simultaneously affecting both the outgoing beam direction
and the direction in which the receiver (APD array) looks. Another
two parameters describe the pointing offset between the outgoing beam
and the receiver direction. A return will only be strong enough to
be readily identified if the well-concentrated beam illuminates the
reflector, the receiver is aligned to collect the returning photons,
and the APD is turned on at the appropriate time. Figure~\ref{fig:pointing}
illustrates five of these six parameters.

The relative alignment between the outgoing beam and the receiver
may not be an obviously variable parameter. But one must intentionally
point ahead of the lunar reflector---to where it \emph{will} be in
1.25 seconds---while looking behind the ``current'' position of the
reflector---to where it \emph{was} 1.25 seconds ago. At the transverse
velocity of the moon ($\sim1,000$~m/s), this translates to approximately
1.4~arcsec of intentional misalignment between transmitter and receiver.
But because the telescope mount is driven on altitude and azimuth
axes, the offset direction rotates relative to the instrument depending
on where the moon is in the sky. Additionally, the earth rotation
($\sim400$~m/s) changes the magnitude of the necessary offset. Because
these effects are comparable in magnitude to the divergence of the
beam ($\sim1$~arcsec) and to the field of view of the receiver (1.4~arcsec),
they must be accommodated in an adjustable manner.

Acquisition and alignment are both greatly aided by the CCD camera (SBIG
STV model), which picks off 7\% of the incoming light at the front of the
receiver tube. Because there are no moving optics between the CCD camera
and the APD array, a particular pixel on the CCD may be identified with the
center of the APD field. It is therefore straightforward to ensure receiver
alignment to a star, or to a crater or other identifiable lunar landmark. A
narrow-band notch filter in an actuated filter-wheel within the CCD
prevents green laser light from saturating the CCD so that we may continue
to view the target while the laser is flashing.

The CCD is also used to verify the transmitter pointing direction
by looking at the return from the fiducial corner cube on the CCD.
This is done with the laser turned to very low power, the T/R optic
in a transmissive orientation, and an uncoated piece of glass replacing
the notch filter within the CCD. The $\sim4$~arcsec diffraction
spot from the corner cube is easily visible on the CCD camera, and
its position compared to the known APD position. When the transmit
beam and receiver are precisely co-aligned, the corner cube diffraction
spot appears centered on the APD position. The relative transmit/receive
alignment is adjusted via M7, which is in a tip-tilt actuated mirror
mount. The intentional transmit/receive offset angle is likewise confirmed
using this system. But because we perform the fiducial alignment check
with the T/R optic in a clear position, its phase is rotated relative
to the actual transmit phase during ranging. Any misalignment between
the surface normal and the rotation axis manifests itself as a wobble
in the T/R optic, potentially invalidating the alignment. We can map
this wobble in a straightforward manner---just looking at the apparent
fiducial return position as a function of T/R optic phase---and correct
for it. Indeed, we must offset the beam by two arcseconds when the
T/R optic phase is in the nominal clear position to accommodate this
misalignment.

Divergence of the beam is checked in two ways. First, a shear plate
in a section of collimated beam ensures that the wavefront has a very
large radius of curvature (e.g., near planar). This is sensitive at
a level corresponding to 0.05~arcsec of divergence outside the telescope.
To the extent that the telescope---when properly focused---delivers
a collimated beam after L3, this measurement ensures that the reverse
path (collimated beam sent into L3) will deliver a collimated beam
at the output of the telescope. Because the CCD and associated lens,
and the APD together with lens L3, have both been set to focus at
infinity, the apparent best-focus as determined by either device achieves
proper collimation of the output beam. In the second method, a corner
cube is inserted at various points around the periphery of the outgoing
beam. The position of the corner cube return on the CCD should not
change as a function of which side of the beam the corner cube is
placed. This test is sensitive at the $\sim0.5$~arcsec level---comparable
to the precision with which telescope focus is determined in realistic
scenarios.

The gate timing is based on predictions generated from the Jet Propulsion
Laboratory's DE403 lunar ephemeris. Appropriate corrections for relativistic
time delay, atmospheric propagation delay, polar wander, and earth
rotation are incorporated. For the duration of the run, the predicted
round-trip travel-time to each of the four available reflectors is
calculated at five minute intervals. A polynomial of order 8--12 is computed
to fit the predictions to roughly sub-picosecond precision. The time
of laser fire is latched to sub-microsecond precision, and used to
compute the polynomial delay. Thus far, we have always seen the return
signal arrive within 10~ns of the prediction, and suspect that we
can refine our software and site coordinates to achieve 1~ns precision.

If the timing, beam divergence, and transmit/receive offset are properly
established, the only remaining uncertainty is the telescope pointing.
The APOLLO lunar pointing model is based on the selenographic coordinates
of each feature, with libration, parallax, etc. all computed at the
time the telescope track command is issued. Thus, once corrective
pointing offsets are determined for one feature, any other feature
may be acquired with little pointing error if the corrective offset
is maintained from one pointing to the next. We can typically make
moves across the moon's face without accumulating more than one arcsecond
of pointing error. Because none of the reflectors are directly visible
from the ground, we must rely on local topographical features whose
centers are well-defined. Near each reflector (within about 100 arcseconds)
we have identified reference features that are compact and visible
at all solar angles. When acquiring the reflector, we center on the
reference crater, then issue the command to track the reflector. We
often find the return signal at this nominal pointing position, though
it is straightforward to execute a 1~arcsec spiral raster pattern
around the nominal position to pick up the signal. Once the signal
is established, we may try to optimize the telescope pointing, the
transmit/receive offset, and telescope focus.

\section{Mechanical Considerations\label{sec:Mechanical-Considerations}}

The laser is built onto a $\sim100$~mm-thick honeycomb optical bench,
occupying a $610\times1220$~mm$^{2}$ section of the $915\times1220$~mm$^{2}$
bench. The T/R optical switch, receiver tube, and CCD camera are all
mounted to the remaining section of the bench---a $305\times1220$~mm$^{2}$
strip beside the laser (Figure~\ref{fig:optics}). Thus the transmitter
and receiver share the same rigid platform, so that their relative
alignment (and intentional offset) is preserved at all telescope orientations.
Any flexure of the optical bench relative to the telescope would appear
as a pointing error in the telescope. The optical bench is affixed
to the telescope via a system of six flexures, each stiff in one dimension
(for both tension and compression) and flexible in the other two.
The flexures are arranged to critically constrain the optical bench
relative to the telescope structure in its six degrees of translational
and rotational freedom. These six flexures contact the telescope at
three locations in a triangular pattern, each near structural webs
within the steel primary mirror cell frame. The purpose of the flexures
is to allow thermal expansion between the thermally stabilized optical
bench and the ambient-exposed primary mirror cell. In extreme conditions,
this may reach one millimeter of differential contraction. The flexure
system accommodates this motion without imposing stresses on the optical
bench.

\section{Electronics Implementation\label{sec:Electronics-Implementation}}

The most important electronics in the APOLLO system are those responsible
for the time measurement corresponding to the lunar range. Other systems
monitor temperatures, fluid flow rates, laser power, laser pulse energy,
telescope tilt, and actuate optics, adjust laser setpoints, and control
power to the various devices in the system. Another system provides
interlock control of the laser shutter for aircraft avoidance and in-dome
safety.  This section concentrates on the timing system implementation, as
the rest---while important for the operation of the system---is not
critical to the scientific quality of the data.

\subsection{Detector Electronics\label{sub:Detector-Electronics}}

\begin{figure}
\begin{center}\includegraphics[%
  scale=0.75,
  angle=0]{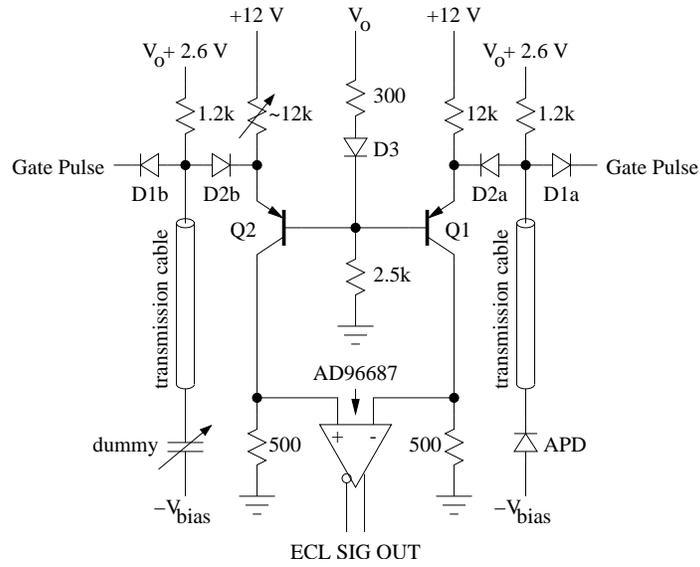}\end{center}

\caption{APD ``daughter board'' electronic readout scheme. The APD is
biased with $-V_{\mathrm{bias}}\approx-24$~V, about 1~V short of
breakdown. $V_{0}$ is typically about 5~V, and controls how much
excess voltage is applied to the APD cathode when the gate pulse is
brought to a positive voltage (about 7~V). The APD is mirrored by
a ``dummy'' capacitor acting like a non-avalanching APD. A fast
comparator senses current draw of the APD when it undergoes avalanche.
\label{fig:APD_elec}}
\end{figure}

The $4\times4$ detector array is packaged in a 40-pin dual-inline
package, which is situated in a socket on a multi-layer electronics
board. Directly behind the APD is an array of variable capacitors
whose capacitance (0.4--2.5~pF) roughly matches that expected from
the APD elements themselves. Each capacitor and APD element is connected
via low-capacitance coaxial cable to an individual, modular ``daughter''
board inserted into one of 16 sockets on the main board. Figure~\ref{fig:APD_elec}
shows the differential APD triggering electronics as implemented on
the daughter boards. The scheme closely follows that presented in
\citet{apd_scheme}.

In the ``gated off'' state, the ``gate pulse'' voltage is
held about one diode-drop below ground, so that diodes D1a and D1b
conduct (D2a and D2b are reverse-biased), and voltage applied to the
APD cathode and to the ``dummy'' capacitor is $\sim0.0$~V. The
APD anode is held at $-24$~V, which is just below the breakdown
voltage of the device. In the ``gated on'' state, the gate pulse
is brought to $\sim7$~V, so that diodes D2a and D2b conduct (D1a,
D1b are reverse-biased) so that a positive voltage near $V_{0}\approx5$~V
is applied to both the APD cathode and the matching capacitor. The
total voltage seen by each APD element (and capacitor) exceeds the
breakdown voltage, placing the APD into avalanche Geiger mode.

The readout scheme directs current from the APD/capacitor across 500~$\Omega$
resistors isolated from the APD/capacitor by the transistor. The voltage
increase on the sensing resistors associated with turning on the gate
is common to both comparator inputs. A single photon entering the
APD causes it to avalanche---an action not mimicked by the capacitor---reducing
the current through Q1 and triggering the fast comparator (AD96687)
to produce an ECL (emitter-coupled logic) output signal. The comparator's
output signal is routed from the daughter board through the main board
to the high-resolution time-to-digital converter (Section~\ref{sub:High-resolution-Timing}).
Returning to the ``gated off'' state quenches the avalanche in
preparation for the next event.

\subsection{Timing System\label{sub:Timing-System}}

\begin{figure}
\begin{center}\includegraphics[%
  scale=0.75,
  angle=0]{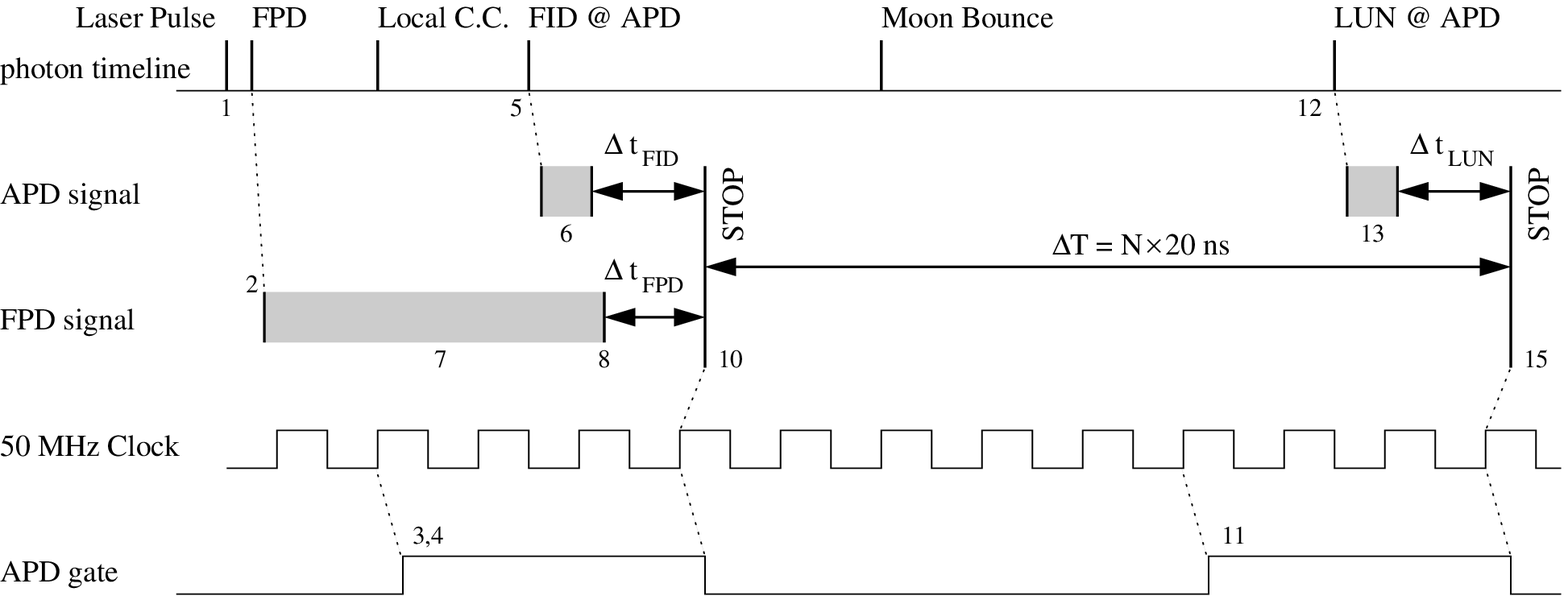}\end{center}

\caption{Sequence of events in the APOLLO measurement. The photon timeline
starts with the laser pulse. A small fraction of the light hits the fast
photodiode to generate a high-precision measurement of launch time,
and another small fraction is intercepted by the local corner cube, to be
sent back for detection by the APD as a fiducial (FID) event. The
light that bounces off the lunar reflector is detected by the APD as a
lunar (LUN) event, the APD being gated on for $< 200$~ns around the time of
arrival of each event. Causally related events are indicated by dotted
lines, and cable delays are indicated by shaded regions. The
time-to-digital converter (TDC) measures the $\Delta t$ values indicated in
the figure to 25~ps resolution, referenced to a STOP pulse sliced from a
low-noise 50~MHz clock pulse train. Counters track the number of clock
pulses, $N$ (8 in this case), between STOP pulses, measuring $\Delta T$. The
fundamental measurement is the round-trip travel time from the local corner
cube to the lunar reflector and back, given by $\tau=\Delta T+\Delta
t_{\mathrm{FID}}-\Delta t_{\mathrm{LUN}}$.  The time axis is not to
scale.  Numbered events correspond to numbered signals in
Figure~\ref{fig:timing}.\label{fig:sequence}}
\end{figure}

\begin{figure}
\begin{center}\includegraphics[%
  scale=0.75,
  angle=0]{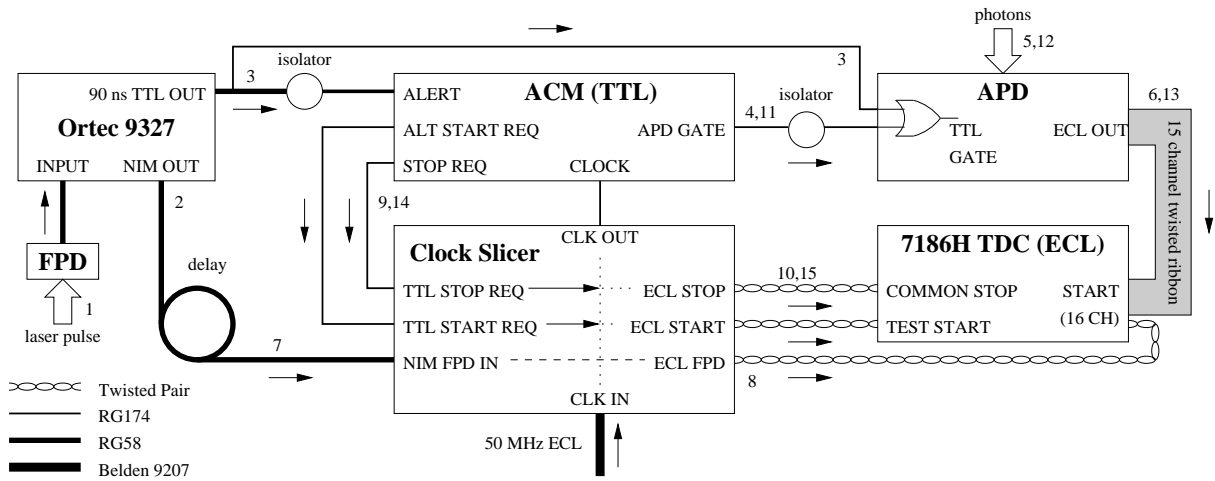}\end{center}

\caption{The core elements of the timing system. The ACM, TDC, and Clock
Slicer all reside in the CAMAC crate in the "Utah" box on the telescope.
The STOP and START pulses out of the Clock Slicer are derived directly from
the high-quality 50~MHz clock source, latch-enabled by the TTL requests
from the ACM. One input channel of the TDC is used for the appropriately
delayed FPD ``timing anchor'' pulse. The TTL output of the Ortec 9327 is
split so that it may both alert the ACM to a laser pulse and initiate the
APD gate. Some signals are ground-isolated by the use of isolation
transformers. Numbers relate to the sequence of events for a fiducial/lunar
pair.  For calibration, the TEST input of the TDC fans out to act as an
input for all 16 channels. \label{fig:timing}} \end{figure}

Figure~\ref{fig:sequence} displays the sequence of events in one cycle of
the APOLLO lunar range measurement.  The timing system, in brief, is built from the following hierarchical
pieces: 

\begin{enumerate}
\item a high-accuracy clock and associated high-quality frequency standard
at 50~MHz;
\item a system of counters to track every clock pulse (thus 20~ns resolution);
\item a 12-bit time-to-digital converter (TDC) with 25~ps resolution and
100~ns range.
\end{enumerate}
The following three sections treat each of these stages. Aside from
the clock, the timing electronics are located in a CAMAC (Computer
Automated Measurement And Control: IEEE 583 standard) crate situated
in the optical bench enclosure. Besides the crate controller (WIENER
PCI-CC32), the inhabitants of the CAMAC crate (Figure~\ref{fig:layout})
are:

\begin{enumerate}
\item A custom clock distribution board, referred to as the ``Clock Slicer''
(though with no CAMAC connections);
\item A custom counter and state machine board with CAMAC interface;
\item A commercial TDC (Phillips Scientific 7186H; 16 channels, common STOP)
with CAMAC interface.
\end{enumerate}
Figure~\ref{fig:timing} depicts the core elements of the timing
system and their means of interconnection.

\subsubsection{System Clock\label{sub:System-Clock}}

The clock is an XL-DC model from Symmetricom (formerly TrueTime) that
is GPS-disciplined to maintain time in accordance with atomic
standards to approximately 100~ns. To achieve 1~mm range precision,
one must know absolute time to better than one microsecond, as earth
rotation and lunar orbit velocities are $\sim$400~km~s$^{-1}$ and
$\sim$1000~km~s$^{-1}$, respectively.  In the APOLLO implementation,
the antenna is about 100~m away from the clock, separated by a
single-mode optical fiber with RF converters on either end. A tracking
loop with a slow time constant ($\approx2000$~s) maintains
synchronization of the clock's internal ovenized quartz oscillator by
applying a control voltage. The control voltage is converted from a
digital solution to an analog voltage via a digital-to-analog
converter (DAC). The DAC value is updated every ten seconds and is
accessible at all times via computer interface. A single step of the
DAC corresponds to a frequency change of $1.5\times10^{-11}$,
translating to about 40~ps ($\approx6$~mm) in the 2.5~s round-trip
lunar range measurement.  This step size is large compared with
APOLLO's target precision. As such, we monitor the clock's DAC value
every ten seconds whether actively ranging or not. By fitting a smooth
function to the DAC, unphysical steps in the lunar range may be
compensated. But even without performing this step, the frequency
offset would average away over several-hour timescales.

The direct output of the clock is a 10~MHz 1~V RMS sinusoid into
50~$\Omega$. The harmonic quality of the sinusoid is very good,
with higher harmonics down by at least 30 dBc. The phase noise is
also exceptionally low, with a high-frequency noise floor of $-154$~dBc,
translating to a zero-crossing jitter of 2.5~ps. Over a 2.5~s period,
the zero-crossing jitter is $\sim5$~ps. 

The 10~MHz clock signal is frequency-multiplied by 5 to become a 50
MHz sinusoid.  The multiplication introduces an unavoidable $1+\log5=1.7$
degradation in phase noise so that the edge jitter over 2.5 seconds becomes
$\sim$8~ps. The clock is housed away from the rest of the timing
electronics in a temperature-controlled enclosure.  It was not viable to
place the clock with the rest of the timing electronics on the telescope
due to a pronounced tilt-sensitive clock frequency offset. Initially, the
10~MHz sinusoid was routed directly to the timing electronics crate via an
11~m RG58 coaxial cable. After July 2006, the $5\times$ multiplier was
repackaged to sit next to the clock, now sending a 50~MHz ECL (square)
clock signal via a Belden 9207 shielded twisted pair cable to the
electronics crate. No obvious degradation accompanied the temporary
arrangement, though the decision to route the 50~MHz ECL rather than the
10~MHz sinusoid was largely based on a desire to accomplish greater noise
immunity.

\subsubsection{Clock Slicer and APOLLO Command Module\label{sub:ACM}}

The 50~MHz clock signal arrives at the ``Clock Slicer'' board in the CAMAC
crate, where it is distributed by a low-jitter, low-skew distribution chip
to a number of comparators.  The version of this board used prior to August
2006 also performed the 5$\times$ clock multiplication, and distributed the
50~MHz sine wave via a transformer distribution system. This was more
susceptible to noise than the current distribution scheme. Among the
comparators in the Clock Slicer, one converts the ECL clock into a TTL
square wave, which becomes the timing reference for the APOLLO Command
Module (ACM). The ACM, under the direction of the hardware control
computer,  performs a variety of functions:

\begin{itemize}
\item Counts clock pulses to achieve 20~ns resolution on event times, establishing
the time within each second by reference to the 1 pulse-per-second
signal from the XL-DC clock
\item Generates STOP (and START) requests for the Clock Slicer
\item Responds to the fast photodiode signal from laser fire to initiate
fiducial gate (APD turn-on)
\item Coordinates the APD gate event to catch the returning lunar photons
\item Latches counters at key events (APD gate events)
\item Requests latched time from the XL-DC clock to microsecond precision
corresponding to gate events
\item Coordinates laser fire with respect to the rotating T/R optic
\item Coordinates the diffuser/attenuator disk rotation and phase relative
to the T/R optic
\end{itemize}
Each gate event produces latched values of various 50~MHz counters,
so that for example, the time within the second of each gate is captured
to 20~ns resolution. The latched time actually corresponds to the
\emph{end} of the gate event, and this same end-of-gate event is used
to generate a STOP pulse in the high-resolution TDC (Section~\ref{sub:High-resolution-Timing}).
A free-running 28-bit counter---wrapping each 5.4 seconds---is also
latched at each gate event, and it is upon this counter that the scheduling
of the lunar gate is based, in conjunction with the prediction polynomial
discussed earlier. The time-within-second and free-running counters
provide pulse-count redundancy, which confirm accurate counting with no
exceptions to date.

\subsubsection{High-resolution Timing\label{sub:High-resolution-Timing}}

At the end of each gate event, the ACM sends a 20~ns TTL signal to
the adjacent Clock Slicer board, which unlatches (enables) for exactly
one clock cycle a fast comparator (AD96687) with the 50~MHz ECL clock
as its input. A single ECL clock pulse is thereby sliced out of the
50~MHz clock train, whose edge is a low-jitter derivative of the
clock signal. This pulse serves as a common STOP pulse to the high-resolution
TDC. The START pulses are generated by photon events at the APD detector
array. The TDC generates a 12-bit digital number corresponding to
a time ranging from 10 to 110~ns, corresponding to about 25~ps per
resolution element. The actual jitter of the TDC is somewhat better,
at $<15$~ps RMS. 

A key part of the high-resolution timing system is the fast photodiode
(Hamamatsu G4176 with 20~ps rise-time coupled to a Picosecond Pulse Labs
model 5545 bias tee with 12~ps rise time) that also serves to alert the
system to laser fire. The FPD signal is received by the Ortec 9327 1~GHz
amplifier and timing discriminator. The discriminator uses a
constant-fraction technique to produce a NIM output having $<10$~ps jitter
relative to the input signal. The combination of FPD,
amplifier/discriminator, and TDC have been verified in the laboratory to
respond to a short laser pulse with a timing jitter less than 20~ps.  The
FPD signal therefore provides a high-precision reference to the laser fire
time. While we ultimately rely on the individual photon returns from the
fiducial corner cube to establish a differential range measurement to the
moon, the FPD ``anchor'' provides a low-jitter indication of laser fire
\emph{for every single pulse}.  We assume that the time offset between FPD
detection and fiducial detections at the APD---which involve cable delays
and varying electronics response times and therefore may change with
temperature---is relatively constant over the 5--10 minute timescales of a
ranging run.

\subsubsection{Calibration of the TDC\label{sub:Calibration-of-the}}

A possible non-differential aspect in APOLLO's ranging scheme stems
from nonlinear properties of the TDC. Some function relates digital
output of the TDC to the actual START-STOP time difference. Unless
the TDC is used in precisely the same range for the fiducial and lunar
gates---a condition that is deliberately arranged via our gate timing---imperfect
knowledge of this function will contribute to a systematic error in
the way that event timing is calculated. We know that the linear gain
(e.g., digital counts per nanosecond) varies with temperature by approximately
160 parts per million per $^{\circ}$C. If the times measured by the
TDC for fiducial and lunar events is allowed to be as large as 50~ns,
160~ppm per $^{\circ}$C translates into 8~ps per $^{\circ}$C,
which is about one millimeter of one-way range.

The ACM, together with the Clock Slicer, is capable of establishing
a calibration of the TDC. In much the same way that the ECL STOP pulse
is created by the Clock Slicer at the request of the ACM, the ACM
may also request an ECL START pulse from an earlier 50~MHz clock
transition. The low jitter of the clock signal therefore guarantees
that the separation of the START and STOP pulses is an integer multiple
of 20~ns to $\sim10$~ps precision. In this way, START/STOP pairs
are created at 20, 40, 60, 80, and 100~ns separation. 1,000 pairs
of each type are sent to the TDC at a rate of 1~kHz, so that the
calibration procedure is carried out in about five seconds. The START
pulse is applied to the TEST input of the TDC, which distributes the
ECL signal simultaneously to all 16 channels of the TDC. The calibration
is carried out at the beginning and end of each ranging sequence,
bracketing the roughly 5--10 minute runs.

In addition to determining the gain, offset, and approximate nonlinearity
of each TDC channel, the calibration procedure provides a means to measure
both the clock jitter (at high frequencies) and the TDC jitter.  For each
START/STOP pair separation (i.e., 1,000 events), each channel will possess
an associated mean of the digital delay-number reported by the TDC. For
each event pair, one may then compute the offset between the reported delay
and the mean delay. If there is a systematic offset (across the 16
channels) from the mean for a given event pair, it may be concluded that
the time-separation of the externally-generated pair was itself
systematically off. In this way, one may separate the distribution of
external offsets from internal TDC jitter. Typical results indicate an
external (clock, multiplier, and Clock Slicer electronics) jitter of 10~ps
and a TDC jitter of 15~ps. The separation cannot be without uncertainty: 16
channels of TDC values at 15~ps jitter may have a random collective offset
from the mean of about 4~ps. But compared to 10~ps in a quadrature sense,
this is a small influence.

\subsubsection{Error Budget\label{sub:Error-Budget}}

Based on measurements in the laboratory of the timing performance
of the individual components comprising the APOLLO system, we derive
the RMS error budget per photon presented in Table~\ref{tab:error_budget}.
The timing errors are presented as round-trip measurement errors,
while the errors expressed in millimeters represent one-way errors
at a conversion of 0.15~mm~ps$^{-1}$. The APOLLO system, according
to this tabulation, has a system random uncertainty of 93~ps, or
about 14~mm. In this case, the typical uncertainty arising from the
tilted retroreflector array dominates the error budget, so that the
required photon number is effectively determined by the lunar libration
angle, and ranges from about $20^{2}=400$ to about $47^{2}\approx2200$.

\input{tab1}

In practice, the APOLLO system does not achieve this performance,
most likely due to electromagnetic noise present at the time of laser
fire. Two tests have supported this statement. First, two fast-photodiode
units coupled to Ortec 9327 discriminators were triggered from the
same laser pulse, and their times as registered on the TDC unit compared.
In the laboratory, using a diode pulse laser, this measurement yields
a 20-25~ps RMS error in the comparative timing. In the implementation
at the telescope, the same test produces a comparative timing with
$\sim55$~ps RMS variation. Second, we can perform the calibration
tests discussed in Section~\ref{sub:Calibration-of-the} synchronized
to the laser fire, such that any impact of laser noise on the Clock
Slicer and TDC units can be seen by comparing to the performance when
the laser is not firing. These tests show an intrinsic TDC jitter
going from 15~ps to 25~ps, and jitter on the external signals (arriving
into the TDC) going from $<10$~ps to 30~ps.

In looking at the spread of the fiducial corner cube measurement---which
should come out near the 93~ps estimate of Table~\ref{tab:error_budget}---we
find typical RMS errors of 180~ps and 168~ps with and without the
diffuser in place, respectively. Because these numbers are considerably
larger than the observed degradation of the TDC and fast-photodiode
measurements discussed above, we conclude that the electromagnetic
interference from the laser has a significant
influence on the timing measurement of the APD signals. At present,
APOLLO is still usually dominated by the reflector tilt, even though the
system is not as precise as intended.

Most of the electromagnetic noise from the laser stems from the Marx
bank generating a $\sim4000$~V pulse with a $\sim1$~ns rise time
to dump the pulse out of the oscillator. We are in the process of
designing a shield for the Marx bank and associated Pockels cell such
that the light signal that triggers the switch is fed into the shield
via optical fiber, and only heavily filtered DC power lines penetrate
the shield. Initial laboratory tests indicate that substantial shielding
is possible, so that APOLLO's intrinsic per-photon error budget may
approach the design goal in the near future.

\section{Thermal Control and Monitoring\label{sec:Thermal-Control}}

Despite the differential measurement mode employed by APOLLO, thermal
control is important. The differential measurement technique assumes
that the system performance has not changed during the 2.5 second
round-trip travel time of the pulse. But to take maximum advantage
of the timing ``anchor'' discussed in Section~\ref{sub:Requirements-Differential},
we want the system to be stable for longer periods of time---over
several minutes. If we tune the fiducial return to be about one photon
per pulse, and allow a single-photon measurement uncertainty of 100~ps,
reaching a 5~ps goal requires 400 photons in each channel. This translates
to 6000 photons across the array, and thus about 6000 shots, taking
5 minutes.

The chief systematic thermal vulnerability in APOLLO is the variation
of gain (picoseconds per bin) in the TDC as a function of temperature.
At roughly 160 parts per million per $^{\circ}$C, a measurement in
the middle of the TDC range (50~ns) translates to 8~ps per $^{\circ}$C,
which corresponds to nearly one millimeter of one-way range per $^{\circ}$C.
One solution is to arrange the timing of the lunar gate such that
the TDC measurements span the same range for both fiducial and lunar
photons. Doing this to even 5~ns precision reduces thermal coupling
to the level of 0.5~ps per $^{\circ}$C---a tolerable level.

All the same, it is desirable to regulate the temperature of the apparatus.
Among other things, this promotes stability in the operation of the laser.
Our goal, then, is to regulate the thermal environment at the
$\sim1\,^{\circ}$C level. We have deployed surface-mount resistive
temperature devices (RTDs) throughout the APOLLO apparatus to monitor the
thermal state. The hardware control computer periodically reads the
temperatures and effects thermal controls to maintain temperature. Because
the thermal state of the TDC is especially important, we have five RTDs
distributed within the TDC, arranged in a vertical column at the positions
of the charging capacitors whose charge is proportional to the measured
time interval. The gradient within the device allows us to interpolate to
intermediate channels for an estimate of temperature channel-by-channel.

In addition to the desire to maintain a steady thermal state, the observatory
requires us keep thermal emission into the telescope
enclosure below 50~W, to avoid creating local turbulence along
the telescope beam path that affects image quality (both for ourselves
and for subsequent observers). The result is a thickly insulated laser
enclosure employing polyisocyanurate at a thermal conductivity of
0.02~W~m$^{-1}$~$^{\circ}\mathrm{K}^{-1}$ and a thickness of
8.5~cm. This enclosure maintains a temperature difference, $\Delta T$,
at a power loss of $\approx1.5\Delta T$ W. The enclosure is maintained
at 20--21~$^{\circ}$C. A 100~W heater maintains states of positive
$\Delta T$ when the system is off, and a closed-cycle chiller coupled
to heat exchangers within the enclosure maintains the thermal state
when $\Delta T$ is negative, and also when the equipment is on ($\sim250$~W
total power).

\section{Software Control\label{sec:Software-Control}}

The overall control scheme consists of a hardware control computer (HCC),
an instrument control computer (ICC), and a telescope-user interface
\citep[TUI:][]{tui}. The TUI is a highly capable and versatile platform
developed for the operation and control of telescopes at APO. The TUI was
specifically engineered to be augmented for control of various observatory
instruments, and APOLLO utilized this framework for its interface. The TUI
expects to communicate with an ICC, but we did not want to burden the
actual HCC with the ICC communications tasks, so placed a separate ICC
machine between the TUI and the HCC. The ICC is then free to perform
additional tasks including web service and video streaming without
compromising the near real-time performance of the HCC.

\subsection{Hardware Control\label{sub:Hardware-Control}}

The HCC is implemented as a Pentium~III machine running the Redhat
7.2 distribution of Linux. This machine hosts CAMAC, GPIB, and RS-232
serial interfaces, and incorporates a National Instruments 6031E 64-channel
16-bit analog-to-digital converter and 8-channel digital I/O card.
The CAMAC interface is responsible for communication with the TDC
and the ACM. The GPIB interface communicates with the XL-DC GPS-disciplined
clock and the New Focus 8731 optics actuation controller. Serial interfaces
are used to command the laser electronics rack, control the T/R optic
drive motor, read the laser power meter, and interface to the SBIG
STV CCD controller. Analog inputs are used to read the RTD temperatures,
flow-meters and flow alarms, laser pulse energy, and telescope tilt-meter.
The 8 digital outputs are used to activate a configurable set of power
relays for turning equipment on and off. A terminal server provides
additional RS-232 interfaces to the chillers, a few programmable resistors
for remote actuation of potentiometer ``knobs,'' and also provides
control of additional power relays for device activation.

Because the HCC has access to all temperature and flow data, as well
as command of the power states of the various APOLLO devices, the
HCC controls the temperature of the system. This
allows a ``smart'' approach to thermal control: upper and lower
setpoints; programmable hysteresis, and anticipatory action when powering
up for a run. Also running in the background is a check on the XL-DC
oscillator statistics---updated every ten seconds. Tracking this information
allows a reconstruction of the discrete steps in frequency introduced
to keep the average oscillator rate in agreement with the GPS time
reference. All such background data appends a log that is archived
and renewed daily.

The primary function of the HCC is to coordinate ranging activity.
Various operational states are defined in the HCC control software.
For example, in the RUN state, the HCC commands the T/R mirror to
spin-up, performs the TDC calibration, then configures the ACM to
flash the laser according to the T/R motor's encoder/index pulses.
Once the laser is firing, CAMAC interrupts alert the HCC that new
data is available, at which point the TDC and ACM counter values are
read, the pulse energy is read, and the gate time for the associated
lunar return is queued. On a CAMAC interrupt associated with a lunar
event, the TDC and ACM values are recorded, and the next queued gate
time is deployed to the ACM. The primary timing events thus are read
at a 40~Hz rate. Slower and less time-critical activities---such
as serial and GPIB communications to devices---are threaded so that
they do not block the primary activity. At the end of the RUN sequence---usually
a fixed number of laser shots---the HCC performs a final TDC calibration.

Other states are defined to warm-up or cool-down the apparatus, perform
in-dome ranging simulations, stare at a star with the APD, obtain
dark or flat fields from the APD, calibrate the TDC both with and
without the laser firing, measure the average laser power, etc. Each
state creates a data file containing a summary of the device configuration
and all associated data, including background environmental data that
also populates the log file.

\subsection{User Interface\label{sub:User-Interface}}

The HCC permits users to log in via a local telnet session so that
HCC commands may be entered directly. In addition, the ICC may establish
a similar connection, passing information from the HCC to the TUI
session(s), and also passing commands from the TUI to the HCC. Multiple
TUI sessions may be active at any given time, allowing APOLLO participants
to control/monitor the full instrument activity from any location
having a TCP/IP connection. A Python-based APOLLO extension to TUI
provides a graphical interface to the APOLLO apparatus, with tabbed
control over basic HCC operation, the STV CCD camera, the laser, device
power, and a lunar pointer tool.

The APOLLO TUI extension employs HippoDraw to graph incoming data in real
time, such as histograms, stripcharts, APD spatial response, etc.  The
HippoDraw plots become our primary feedback for signal acquisition and
optimization.  The user interface, augmented by streaming video from the
CCD camera, permits operation of the apparatus at any remote location with
high-speed Internet access.

\section{Performance Summary\label{sec:Performance-Summary}}

\begin{figure}
\begin{center}\includegraphics[%
  scale=0.75]{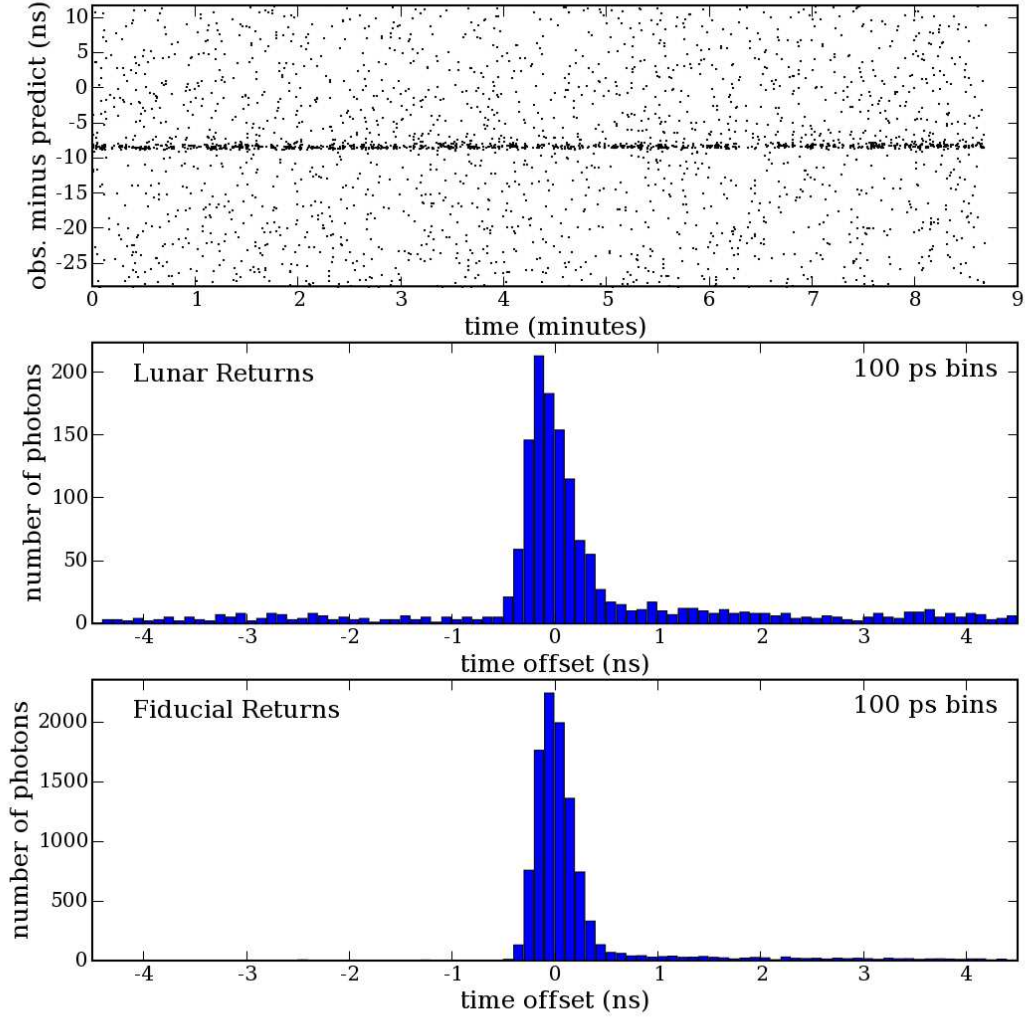}\end{center}

\caption{Example return from the Apollo 11 array on 03 December 2006, in which
1087 photons were collected in a 10000 shot run. The reflector return
is clearly visible against the background. The asymmetry in the background---seen
in both lunar and fiducial histograms---results from diffusion of
some photoelectrons in the APD. The estimated one-sigma error in
determining the lunar range from this run is 1.2~mm.\label{fig:apollo11}}
\end{figure}

In the summer of 2005, small amounts of telescope time were made available
for system engineering and shakedown, during which we made several attempts
to detect a return signal.  The instrument was completed with the
installation of the microlens array in October 2005, just before the first
scheduled telescope time for APOLLO.  In this first observation period in
October, we achieved record returns---the best of which garnered $\sim675$
photons in 5000-shots (250 seconds) for a 0.135 photon-per-shot average. In
subsequent months, we saw sustained rates of 0.25 photons per shot,
occasionally peaking (for 15 second periods) at 0.6 photons per shot. These
numbers refer to the Apollo~15 array, which is three times larger than the
Apollo~11 and Apollo~14 arrays. Figure~\ref{fig:apollo11} shows an example
return from the Apollo~11 array.

Comparison of the lunar return to the fiducial return in Figure~\ref{fig:apollo11}
shows that the fiducials are nearly background-free relative to the
lunar return, both share the asymmetric tail, and the fiducial return
is slightly narrower than the lunar return. The difference in background
is due to the substantial attenuation employed for the fiducial return,
blocking any background light from the moon. The asymmetric tail is
due to photoelectrons created beyond the depletion region in the Geiger-mode
APD, discussed in Section~\ref{sub:Detector-Array}. The difference
in temporal width is a result of the tilted lunar reflector array. In this
case, the lunar return has a FWHM of about 470~ps, while the fiducial
width is about 390~ps. The inferred array tilt introduces a $\sim$265~ps
FWHM, which is consistent with expectations for the libration tilt of
5.8~degrees on that night.  Using a correlation technique to compare
the timing between fiducial and lunar returns results in a 
1.2~mm RMS range uncertainty for this run.

The return photon count rate has a high degree of shot-to-shot variability
due to speckle structure and other ``seeing'' effects imposed on the
outgoing and return beams. For example, in a 10,000-shot run on Apollo~15
where no changes to telescope pointing or device configuration were made,
1239 single-photon events were recorded, 319 two-photon events, 79, 29, 8,
4, 1, and 1 events with 3--8 photons, respectively, for a total of 2303
photons. Thus 46\% of the returning photons were in multi-photon bundles.
Strict binomial statistics at an average rate of 0.23 photons per pulse
would suggest that $<20$\% of the returns would be in multiple-photon
bundles, with no events having more than four photons in a 10000-shot run.

APOLLO is highly sensitive to atmospheric seeing conditions. Not only
does the flux placed on the lunar reflector scale as the seeing-limited
divergence angle squared, but the 1.4~arcsec receiver also begins
to lose flux when the seeing is significantly worse than 1~arcsec.
For large values of the seeing parameter, the total system response
scales as the seeing parameter to the fourth power. Thus 3~arcsec
seeing is 81 times harder than 1~arcsec seeing. Operationally, we
find it too difficult to work when the seeing is worse than about
2.5~arcsec. Zenith seeing at APO exceeds this less than
10\% of the time, though at high zenith angles where we must often work,
the fraction of unusable time is somewhat higher..

Though we have not seen returns at the rate anticipated by a simple
link budget \citep[possible degradation of lunar reflectors: see][]{weak_LLR},
a rate of 0.25 photons per pulse provides sufficient statistics for
one-millimeter range precision on timescales less than ten minutes.

\appendix

\section{Acronyms}\label{app:acronyms}

The following is a list of acronyms frequently appearing in the text.

\begin{lyxlist}{00.00.0000}
\item [ACM]APOLLO Command Module
\item [APD]Avalanche Photodiode
\item [APO]Apache Point Observatory
\item [APOLLO]Apache Point Observatory Lunar Laser Ranging Operation
\item [AR]Anti-Reflection
\item [CAMAC]Computer Automated Measurement And Control
\item [CCD]Charge-Coupled Device
\item [ECL]Emitter-Coupled Logic
\item [FPD]Fast Photodiode
\item [GPS]Global Positioning System
\item [LLR]Lunar Laser Ranging
\item [Nd:YAG]Neodymium-doped Yttrium-Aluminum-Garnet
\item [RTD]Resistive Temperature Device
\item [TDC]Time-to-Digital Converter
\item [T/R]Transmit/Receive
\item [TTL]Transistor-Transistor Logic
\item [TUI]Telescope User Interface
\end{lyxlist}

\acknowledgments

We are indebted to many people for the success of the APOLLO project.
Ed Turner and Suzanne Hawley---successive directors of the
Astrophysical Research Consortium (ARC)---have generously allocated
director's discretionary observing time to the project on a continuing
basis, and have also facilitated observatory support for our
installation and operation procedures. Brian Aull, Bernie Kosicki,
Richard Marino and Robert Reich of MIT Lincoln Lab contributed in a
crucial manner by allowing us to characterize and use their APD array
technology for APOLLO.  Bruce Gillespie of the Apache Point
Observatory (APO) facilitated APOLLO's interface with the observatory
and pioneered our interface with Space Command, the Federal Aviation
Administration, and local military authorities. Mark Klaene of APO was
instrumental in coordinating site activities and suggesting approaches
to APOLLO instrumentation that would achieve both project and
observatory goals. Other APO staff, especially Jon Brinkmann, Jon
Davis, Craig Loomis, Fritz Stauffer, and Dave Woods have been helpful
in getting APOLLO off the ground.  Russell Owen wrote the ICC
software, for which we are entirely grateful.  Russet McMillan has
performed many of the APOLLO observations. Jeff Morgan assisted with
early project definition and telescope interface issues. Sterling
Fisher helped explore and characterize early design ideas. Many
undergraduate students have assisted our development: chiefly Jesse
Angle, who developed the T/R motor control and Evan Million, who
produced our prediction software.  Daniel Miller, Justin Ryser, and
Aimee Vu also contributed to electronic, software,  and hardware
development, respectively. Jonathon Driscoll at UCSD performed some
timing experiments and further developed the T/R motor interface.  Tim
van Wechel at UW mastered the implementation of the ACM, built by
Allan Myers. Allen White at UCSD contributed substantially to APOLLO
electronics development, with critical help from George Kassabian and
Mike Rezin---especially on the APD electronics and the clock
multiplier and Clock Slicer.  James MacArthur and William Walker at
the Harvard University Instrument Design Lab contributed to
electronics development.  We also thank Mike Vinton of the UW physics
machine shop, and Ken Duff, Tom Maggard, and Dave Malmberg of the
Scripps Institute for Oceanography machine shop for considerable help
in fabrication and design. Jim Williams of the Jet Propulsion
Laboratory, and Randy Ricklefs, Judit Ries, Pete Shelus, and Jerry
Wiant of the University of Texas at Austin all provided valuable
advice about the LLR technique, as did Eric Silverberg.  Jim Williams
and Dale Boggs of the Jet Propulsion Laboratory performed frequent
verification of our prediction quality.  Finally, we express gratitude
for the funding sources that enabled APOLLO.  These include NASA
NAG8-1756, NASA NNG04GD48G, and the National Science Foundation
Gravitational Physics program (PHY-0245061), in addition to
discretionary funds from C.W.S. and T.W.M.

\end{document}

%% file: tab1.tex
\begin{deluxetable}{lcc}

\tablewidth{0pt}
\tablecaption{APOLLO random error budget per photon.\label{tab:error_budget}}
\tablehead{
\colhead{Error Source} & \colhead{RMS Error (ps)} & \colhead{RMS Error (mm)}}

\startdata

APD illumination         & 60       & 9 \\
APD intrinsic            & $<50$    & $<7.5$ \\
Laser pulse              & 45       & 7 \\
Timing electronics       & 20       & 3 \\
GPS clock                & 7        & 1 \\
Total APOLLO             & 93       & 14 \\
Retroreflector array     & 100--300 & 15--45 \\
Total random uncertainty & 136--314 & 20--47 \\
\enddata
\end{deluxetable}